\documentclass{article}
\usepackage[english]{babel}
\usepackage{amsmath,amssymb,graphicx,xcolor}

\newcommand{\nospace}{}
\newcommand{\tmaffiliation}[1]{\\ #1}

\newcommand{\tmem}[1]{{\em #1\/}}
\newcommand{\tmop}[1]{\ensuremath{\operatorname{#1}}}
\newcommand{\tmstrong}[1]{\textbf{#1}}
\newcommand{\tmtextbf}[1]{{\bfseries{#1}}}
\newcommand{\tmtextit}[1]{{\itshape{#1}}}
\begin{document}

\title{\tmtextbf{ Conduction Channels of an InAs-Al nanowire Josephson weak
link}}

\author{
  M. F. Goffman, C. Urbina and H. Pothier
  \tmaffiliation{Quantronics Group, SPEC, CEA, CNRS, Universit{\'e}
  Paris-Saclay, \\
CEA Saclay, 91191 Gif-sur-Yvette, France}
  \and
  J. Nyg{\r a}rd, C. M. Marcus and P. Krogstrup
  \tmaffiliation{Center for Quantum Devices and Station Q Copenhagen, Niels \\ Bohr Institute, University of Copenhagen, Copenhagen, Denmark}
}

\maketitle

\begin{abstract}
  We present a quantitative characterization of an electrically tunable
  Josephson junction defined in an InAs nanowire proximitized by
  an epitax-ially-grown superconducting Al shell. The gate-dependence of the
  number of \ conduction channels and of the set of transmission coefficients
  are extracted from the highly nonlinear current-voltage characteristics.
  Although the transmissions evolve non-monotonically, the number of
  independent channels can be tuned, and configurations with a single
  quasi-ballistic channel achieved.
\end{abstract}

\

\

Superconductor-semiconductor-superconductor weak links are interesting hybrid
structures in which the Josephson coupling energy, and therefore the
supercurrent, can be modulated by an electric field
{\cite{de_lange_realization_2015,larsen_semiconductor-nanowire-based_2015}}.
It is even possible to lower enough the carrier density in the weak link to
achieve the conceptually simple situation of a quantum point contact (QPC), in
which only a small number of conduction channels contribute to transport.
Although these kind of hybrid microstructures have been explored for many
years {\cite{schapers_superconductor/semiconductor_2003}}, inducing strong
superconducting correlations into the semiconductor in a reliable way has been
achieved only recently. A well-defined (``hard'') superconducting gap has been
clearly demonstrated both in InAs nanowires {\cite{changw._hard_2015}} and in
In-GaAs/InAs two-dimensional electron gases {\cite{kjaergaard_quantized_2016}}
by using in-situ epitaxially grown Al contacts. Many experiments
{\cite{albrecht_exponential_2016,deng_majorana_2016,spanton_current-phase_2017,shabani_two-dimensional_2016,van_woerkom_microwave_2016}}
are presently using these hybrid structures because they are promising
candidates to implement topological superconductivity and Majorana bound
states {\cite{lutchyn_majorana_2010,oreg_helical_2010-1}}. A good
understanding of their basic microscopic transport features is therefore
necessary. Here we track the evolution of the conduction channels of a QPC
based on an InAs-Al (core-shell) nanowire {\cite{krogstrup_epitaxy_2015}}, as
gate voltages gradually deplete the weak link region.

Nanowires were dispersed onto a Si substrate covered with 500 nm of silicon
oxide. After an Ar ion milling step ($\tmop{energy} 500 ~ \tmop{eV}$, 90~s,
nominal $\tmop{Al}_2 \text{O}_3$ etch rate $\sim 4 \tmop{nm} / \min$), the Al
shell was contacted by e-beam-evaporated 100~nm-thick micrometer-scale Al
leads. The QPC was then defined by completely removing the Al shell over
150~nm by a selective wet etching step in Transene D. The etching region was
defined by e-beam lithography using a PMMA layer deposited on a few-nm-thick
optical resist that turns the Al-resist interface hydrophobic, hence
preventing the peeling of the whole wire while etching
{\cite{lamarre_positive_1990-1}}. In a subsequent lithography step, Au gates
were fabricated on both sides of the exposed InAs core to allow tuning of the
local carrier density. A micrograph of the device and the schematics of the
measurement setup is depicted in Fig.~1. Symmetric biasing of the junction was
achieved with a bridge of 4 resistances placed on the printed circuit board to
which the sample is wire-bonded. The voltage \tmtextit{V }across the wire is
measured with another pair of leads connected to the bias pads, whereas the
current \tmtextit{I} is deduced from the voltage drop across resistance $R.$
Two independent voltage sources $V_{g 1}$ and $V_{g 2}$ connected to the
side-gates control the depletion of the QPC.

\begin{figure}[h]
\begin{center}
  \resizebox{8.6cm}{!}{\includegraphics{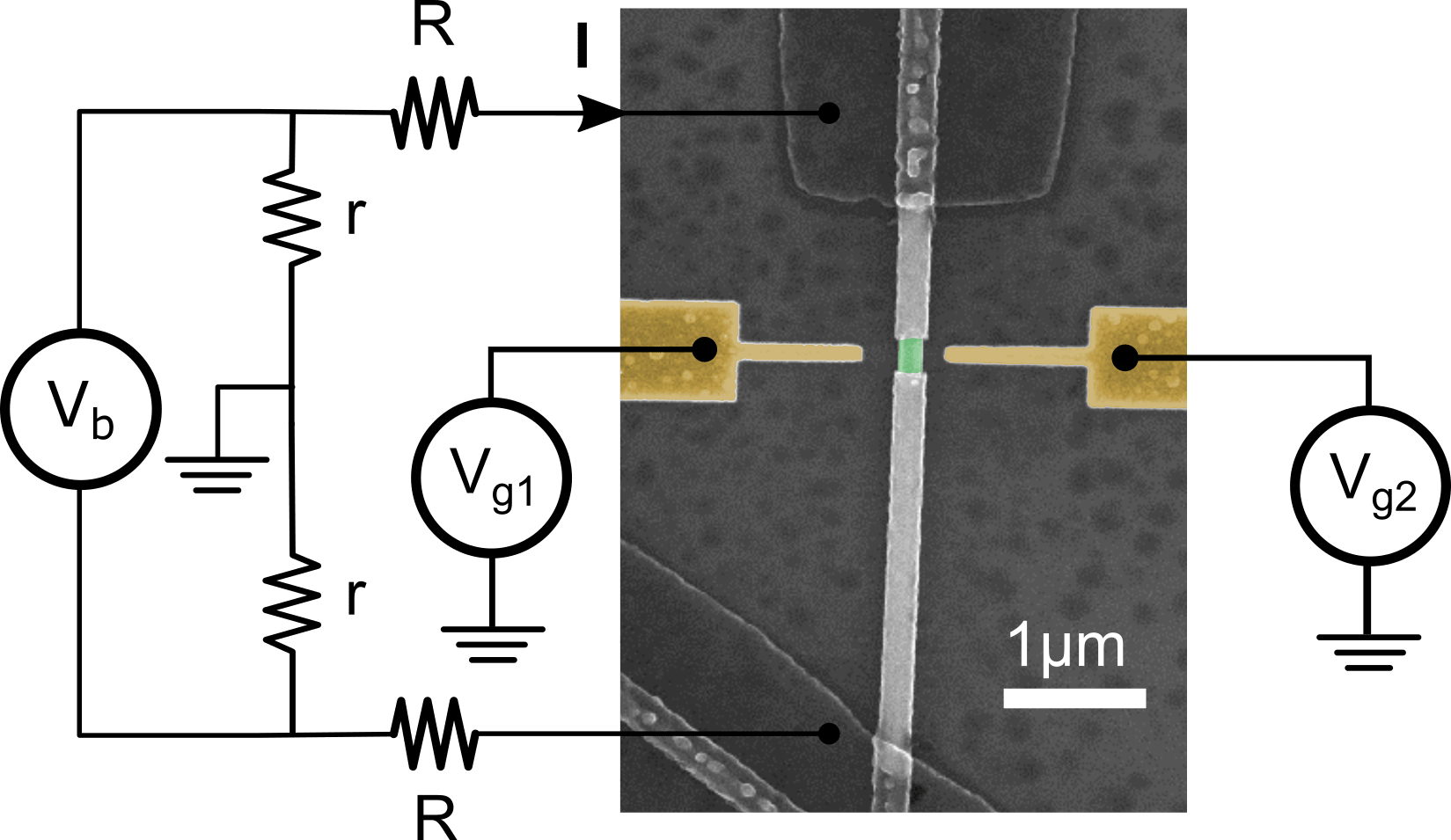}}
  \caption{\label{fig1}Scanning electron micrograph (false colors) of a device
  and schematics of the biasing scheme ($R = 96 \Omega$, $r = 5 \Omega$). An
  epitaxial full-shell nanowire with InAs core (green) and Al shell (grey) is
  connected to micrometer-size Al leads for \tmtextit{I-V} measurements.
  Yellow: Au side-gates.}
\end{center}
\end{figure}

\

\begin{figure}[h]
\begin{center}
  \resizebox{10cm}{!}{\includegraphics{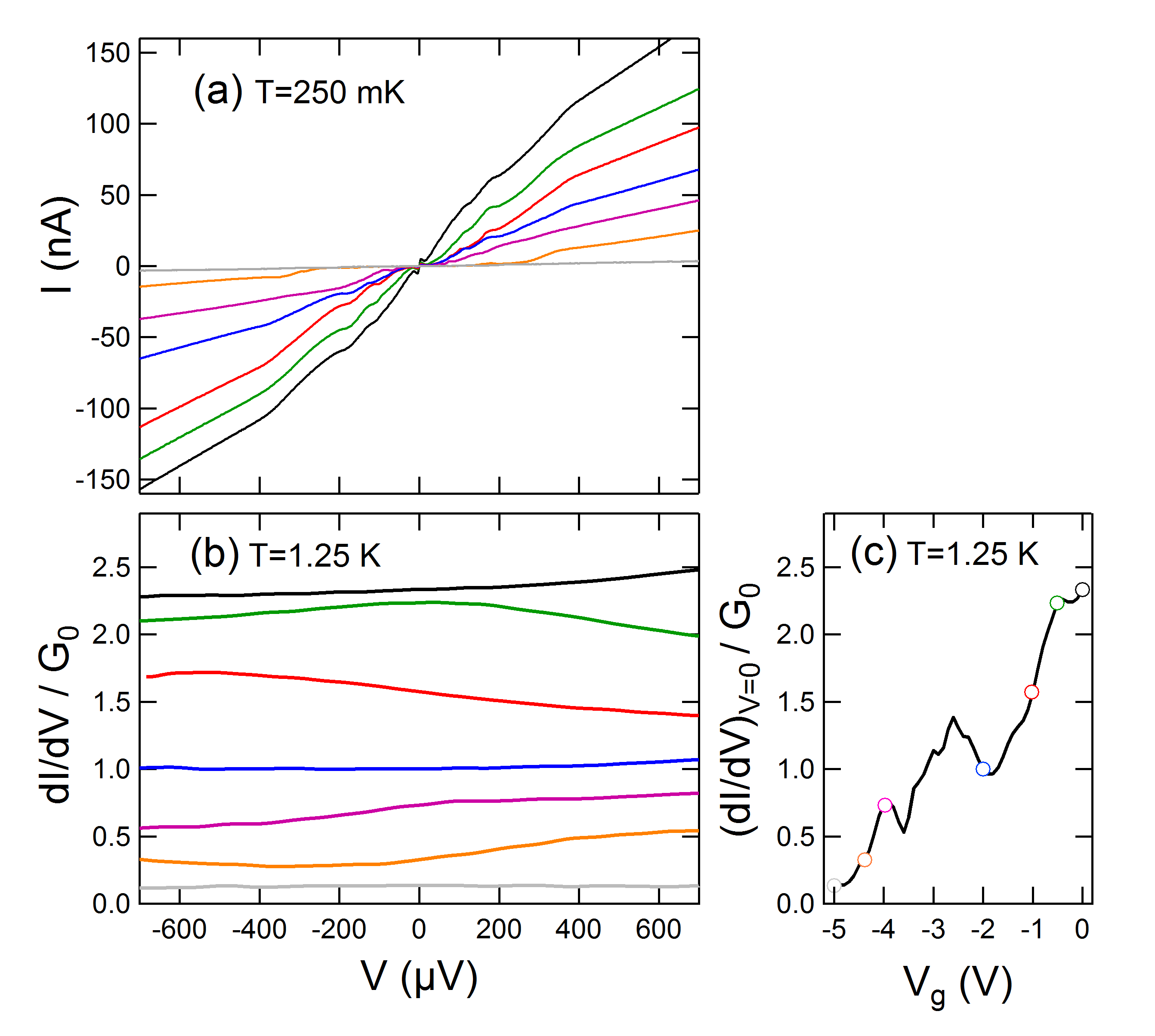}}
  \caption{\label{fig2}(a): \tmtextit{I-V} characteristics measured at 250~mK
  and taken at different gate voltages $V_{g 1}$=$V_{g 2} \equiv V_g $: from
  top (black) to bottom (grey), $V_g =$0, -0.5, -1, -2, -4, -4.4, -5$~
  \text{V}$ respectively. Non-linearities for $| V | \lesssim 300 ~ \mu V$ are
  attributed to multiple Andreev reflections. (b), (c) Normalized differential
  conductance $d \nospace I / d \nospace V / G_0$ in the normal state ($T = 1.25
  \hspace{4pt} \mathrm{K}$), where $G_0 \equiv 2 e^2 / h$ is the conductance quantum; in (b), for the same gate voltages as in (a); in
  (c), as a function of $V_g$, at $V = 0.$ The values of $V_g$ used in (a) and
  (b) are indicated with open symbols.}
\end{center}
\end{figure}

Measurements were carried out in a He3 refrigerator at a base temperature of
250 mK. Figure~\ref{fig2}a shows \tmtextit{I-V} characteristics taken in the
superconducting state at various values of the gate voltages, in the common
mode $V_{g 1}$=$V_{g 2} \equiv V_g$. The overall current decreases as
$V_g$\tmtextit{ }is lowered. This correlates with the reduction of the
differential conductance \tmtextit{dI/dV }in the normal state, as shown in
Fig.~\ref{fig2}b with data at the same values of $V_g $taken above the
superconducting transition temperature of Al. This figure also shows that
\tmtextit{dI/dV} varies with \tmtextit{V}. The complete evolution with $V_g $
of the zero voltage conductance in the normal state is shown in Fig.~2c. The
absence of conductance plateaus merely indicates that the channels do not
close one after the other as $V_g$ decreases, in contrast to what happens with
clean QPCs in two-dimensional electron gases. A detailed and comprehensive
account of the evolution of all transmissions can however be obtained from the
superconducting state data. The \tmtextit{I-V} characteristics in the
superconducting state display kinks at voltages close to $2 \Delta / n
\nospace e$ where $n$ is an integer number \ ($n = 1, 2, 3$ are clearly
visible), $e$ is the electron's charge, and $\Delta \simeq 160 \mu
\mathrm{e{\nospace}V}$ the superconducting gap in the proximitized InAs. These non-linearities result from the charge transport
occurring through Multiple Andreev Reflections (MAR)
{\cite{hurd_current-voltage_1996}}. As shown by experiments on atomic contacts
{\cite{scheer_conduction_1997}}, the number of channels $N$ and the set of
transmissions coefficients $\{ \tau_1, \ldots, \tau_N \}$ can be determined by
decomposing the \tmtextit{I-V} characteristics into the contributions of a few
channels using the well-established theory for MAR
{\cite{averin_ac_1995,cuevas_hamiltonian_1996,shumeiko_scattering_1997}}, with
the gap and the transmissions as adjustable parameters. Note however that the
physics is a bit richer here because, as shown in Fig.~\ref{fig2}b,
$\mathit{\mathrm{}dI/dV}$ varies with $V$ and
\tmtextit{dI/dV{\hspace{1pt}}(V)$\neq$dI/dV{\hspace{1pt}}(-V)}. Although the
MAR theory should then in principle be modified to include the energy
dependence of the transmissions, here we simply treat separately the $V > 0$
and $V < 0$ halves of the \tmtextit{I-V}s and get for each gate voltage value
two estimations of the set of transmissions {\cite{Note}}.

\begin{figure}[h]
\begin{center}
\resizebox{10cm}{!}{\includegraphics{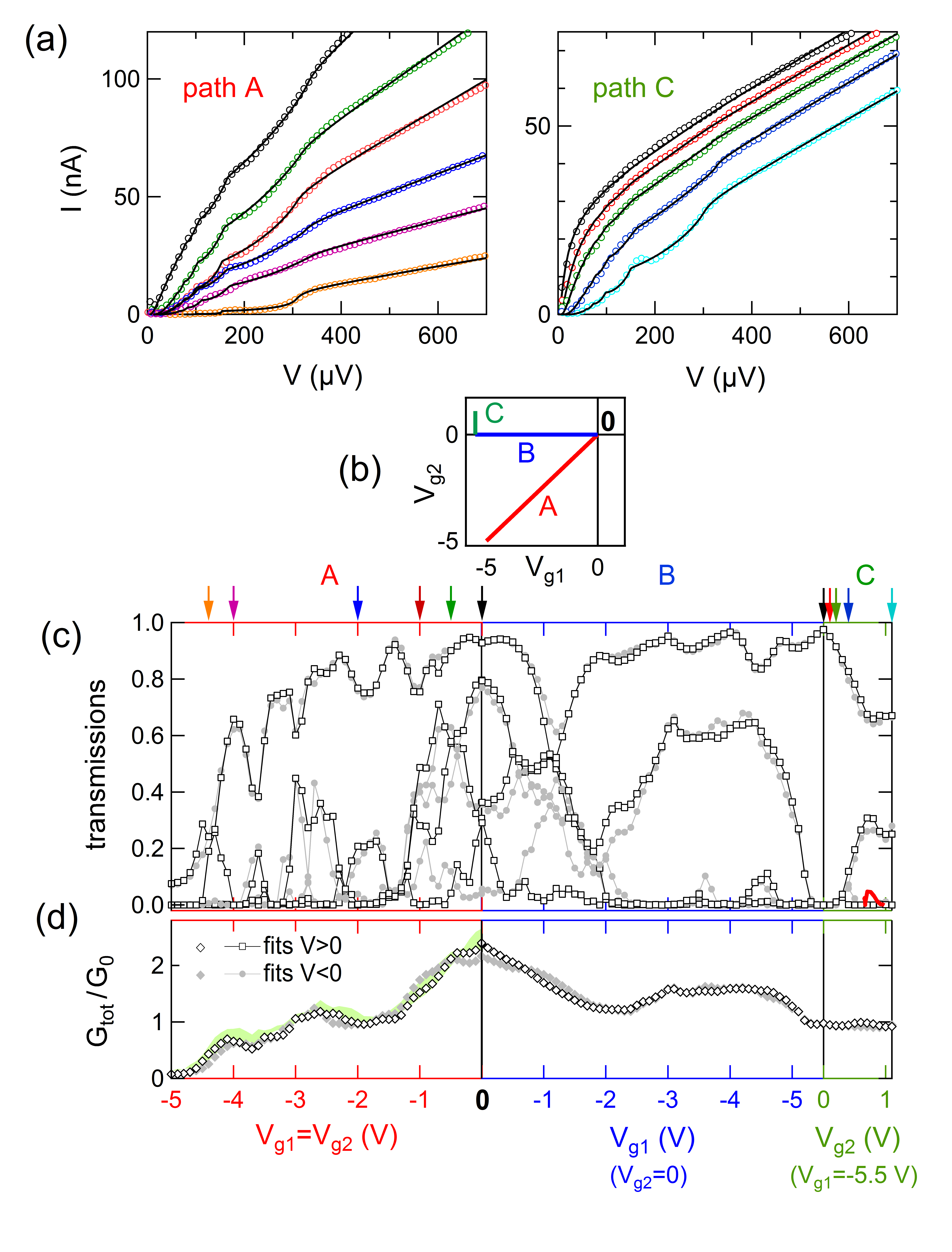}}
    \caption{\label{fig3}(a) Measured \tmtextit{I-V} characteristics (symbols)
    and best fits (lines) considering four independent channels. The curves
    are taken at gate voltages shown with corresponding color arrows in
    leftmost and rightmost panels of (c), following gate paths A and C
    described in (b): path A: sweep $V_g = V_{g 1} = V_{g 2}$; path B: sweep
    $V_{g 1}$ while $V_{g 2} = 0$; path C: sweep $V_{g 2} $while $V_{g 1} = -
    5.5 ~ \text{V} \text{} .$ (c)\&(d) Individual channel transmissions and
    total transmission $G_{\tmop{tot}}$ obtained from the fits as a function
    of gate voltages$.$ Note the inverted gate-voltage scale on the central
    panel that ensures continuity between the 3 plots. Open symbols correspond
    to $V > 0$ fits and full symbols to $V < 0$ fits. In lefmost pannel of
    (d), light green region indicates measured span of normal-state
    differential conductance $d \nospace I / d \nospace V$ for
    the\tmtextit{V}-range over which the fits are performed ($| V | < 640
    \mu$V).}
\end{center}
\end{figure}

The fits, the corresponding channels' transmissions and their sum are shown in
Fig.~\ref{fig3}a, 3c and 3d, respectively. The three panels in
Figs.~\ref{fig3}c\&d correspond to data taken along the three gating paths
shown in Fig.~\ref{fig3}b. The left panel of Fig.~\ref{fig3}a shows the fits
(lines) of the $V > 0$ data of Fig.~\ref{fig2}a (open symbols), which are
taken at gate voltages marked with arrows of the same color on leftmost panel
of Fig.~3c. The right panel of Fig.~3a shows similar data taken at $V_{g 1} =
- 5.5 ~ \text{V},$ for 5 values of $V_{g 2}$ marked with arrows on rightmost
panel of Fig.~3c.

According to the maximum measured normal-state conductance $\mathit{dI/dV}
\simeq 2.5 G_0$, one
expects at least 3 independent conducting channels. Fits are performed using a
Monte-Carlo algorithm described elsewhere {\cite{riquelme_distribution_2005}}.
They require at most 4 channels in the whole gate voltage range investigated.
The mean value obtained for the superconducting gap is $\Delta = 160
\hspace{1pt} \mu \mathrm{e{\nospace}V}$ with a standard deviation of $7 \mu
\mathrm{e{\nospace}V}$. The gap appears to be almost gate-voltage-independent,
as expected from the fact that the Al shell completely covers the InAs regions
that act as superconducting banks for the QPC. \

Good fits with the MAR theory are obtained everywhere except near pinch-off
($V_{g 1}$ and $V_{g 2}$ both below $- 3.5 \mathrm{V}$), where the relative
variations of \tmtextit{dI/dV} in the normal state become large. Furthermore,
in this region, the current is only carried by channels with low transmission,
for which the short junction limit is no longer valid: as argued in
Ref.~{\cite{ingerman_coherent_2001}}, additional resonances develop in the
\tmtextit{I-V} when the channel is longer than $\xi_0 \sqrt{\tau}$, where
$\xi_0$ is the superconducting coherence length in the proximitized InAs.

The evolution of the transmissions as obtained from the $V < 0$
(full circles) and $V > 0$ (open squares)
halves of the \tmtextit{I-V}s is shown in Fig.~\ref{fig3}c.
High transmissions are determined to a good accuracy ($\pm
0.03$), whereas there is some uncertainty on lower transmissions, in
particular when four channels contribute to the current. Despite the fact
that the differential conductance in the normal state depends on $V$, both
determinations give very similar results. Moreover, the extracted total
conductance $G_{\tmop{tot}} = G_0 (\tau_1 + \tau_2 + \tau_3 + \tau_4)$ remains
mostly within the measured range of variation of the differential conductance
in the normal state (see leftmost panel in Fig.~\ref{fig3}d).

When lowering the gate voltages, the number of channels decreases from 4 to 1,
although individual transmissions change non-monotonously. This behaviour is
probably a consequence of impurity charges (unintentional background doping,
deep charge traps in the underneath dielectric, etc.) near the InAs region
giving rise to random fluctuations of the confinement potential in addition to
the one defined by the split gates, as often observed in semiconducting QPCs.
Note that even in the absence of disorder, a Fermi velocity mismatch between
the protected and unprotected regions of the nanowire could also give rise to
standing waves, leading to resonances and non-monotonicity of transmission
with gate voltage.

When only a single gate voltage is decreased (gating path B),
conducting channels evolve more gradually than when both are, as expected.
Interestingly, at $V_{g 1} = - 5.5 ~ \text{V},$ $V_{g 2} = 0$, the current
flows through a single quasi-ballistic channel ($\tau_1 = 0.98$) (black curve
in right panel of Fig.~\ref{fig3}a). One also observes that the {\tmem{I-V}}s
change dramatically along gating path C (right panel of Fig.~\ref{fig3}a),
even though the total conductance (right panel of Fig.~3d) remains almost
constant. This reveals the smooth decrease of the transmission of one channel
compensating almost exactly the increase of the second one (see right panel of
Fig.~3c). Similar transfer of weights are observed at several places in
Fig.~\ref{fig3}c.

\

In summary, we have shown that it is possible to track the conduction
channels' transmissions of an InAs nanowire superconducting QPC by measuring
its current-voltage characteristics. Agreement with the MAR theory in the
short-junction limit is satisfactory except very close to pinch-off. This
gives confidence that following the evolution of the MAR structure with
magnetic field could be a promising method to probe the topological transition
predicted for such InAs nanowires
{\cite{san-jose_multiple_2013,zazunov_low-energy_2016}}. The information
obtained on the transmissions and on the superconducting gap from the
\tmtextit{I-V}s in the superconducting state can be used to better understand
spectroscopy data of the Andreev bound states (ABS) that form when
phase-biasing a QPC. Finally, we find situations where transport occurs
through a single channel of transmission as high as 0.98, leading to ABS
energies of the order of $10 \mathrm{GHz},$which is favorable to probe ABS
using circuit-QED techniques {\cite{janvier_coherent_2015}}.

{\noindent}\tmtextbf{Acknowledgments . }{\tmstrong{}}We thank P. S{\'e}nat and
P.-F. Orfila for technical assistance, A. Levy-Yeyati for discussions and G.
Rubio-Bollinger for supplying us the fitting program. We acknowledge financial
support by ANR contracts MASH, JETS and by the Danish National Research
Foundation.{\hspace*{\fill}}{\medskip}

\

\end{document}